\documentclass[12pt]{article}
\pdfoutput=1
\usepackage[a4paper,left=25mm,right=25mm,top=20mm,bottom=20mm]{geometry}
\usepackage{graphicx}
\usepackage{amsmath}
\usepackage[colorlinks=true,urlcolor=blue]{hyperref}
\usepackage{movie15}

\begin{document}

\begin{center}
{\large \scshape Visualization of a particle's wave function in the double slits experiment}\\
{Eugene B. Postnikov, Artem A. Loktionov}\\
{\it Department of Theoretical Physics, Kursk State University\\
Radishcheva st. 33, Kursk 305000, Russia
}
\end{center}

The double slits experiment is a basic phenomenon, which allows to explain principal behaviour of quantum systems \cite{Feynman,Bohm}. However, textbooks present static pictures of corresponding interference patterns. At the same time, modern computer software for PDE solution provides an opportunity for dynamical modeling of a wave function behaviour using a numerical solution of Schr\"odinger's equation and to use the obtained demonstrations in a teaching of physics. The following material presents such a simulation. 

\vspace{7mm}
In one-dimensional case, Schr\"odinger's equation
\begin{equation}
i\frac{\partial \psi}{\partial t}=\frac{\hbar}{2m}\frac{\partial^2 \psi}{\partial x^2}
\label{Sch}
\end{equation}
has one partial solution \cite{Bohm}
\begin{equation}
\psi(x,t)=\sqrt{\frac{\pi}{\alpha+i\beta t}}e^{i(k_0x-\omega_0t)}e^{-\frac{(x-v_gt)^2}{4(\alpha+i\beta t)}},
\label{wtrain}
\end{equation}
which describes a time-spatial evolution of so called {\it wave packet}, localized structure moving from left to right along x-axis with a group velocity $v_g=\hbar k_0/m$.

{
\centering
\includegraphics[width=\textwidth]{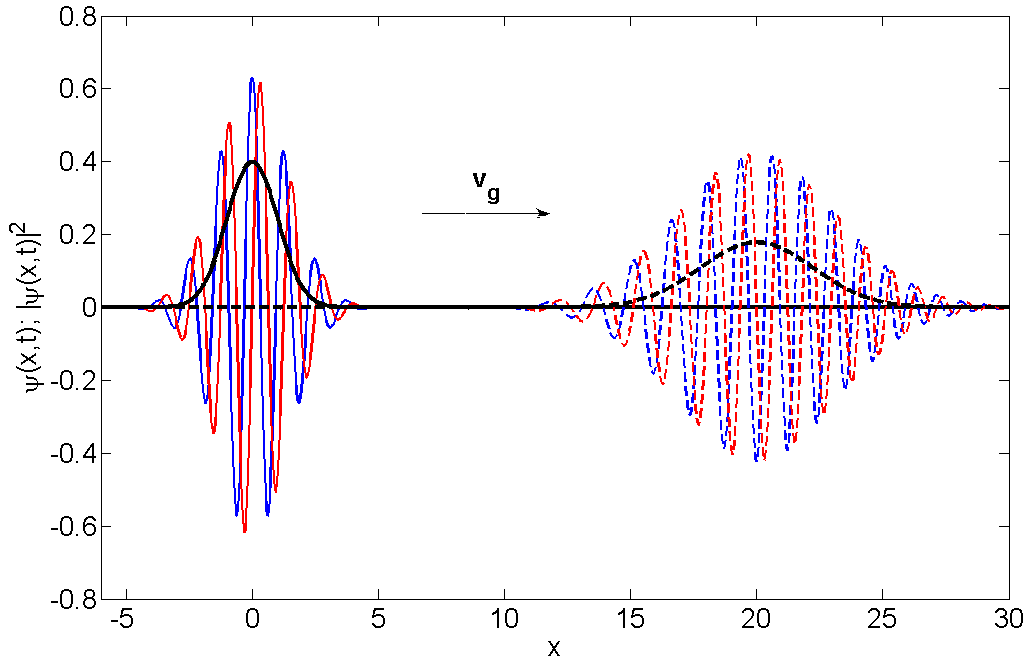}
{\small \sl
Two snapshots of the moving 1D wave packet. Blue and red lines indicate real and imaginary parts of wave function, black corresponds to the
square of its absolute value, i.e. to probability density to detect a particle.
}
}

\vspace{7mm}
The picture above demonstrates, what {\it wave packet} means: real ($\psi_r$, blue curve) and imaginary ($\psi_i$, red curve) parts of a wave function $\psi=\psi_r+i\psi_i$ oscillate while its absolute value squared $|\psi|^2$, which determines the probability density to detect a particle, forms a bell-shaped curve. The latter moves along $x$-axis with velocity $v_g$ that corresponds to the motion of a particle. Thus, a most probable position of the moving particle is determined by the $|\psi|^2$ maximum, and a fast decay of this function indicates that it is well-localized structure, i.e. the probability density to detect this particle has sufficiently non-zero value within a small ``spot''only. At the same time, this wave packet expands with a time ({\it dispersion} of a wave packet). In other words, a width of the bell-shaped curve increases.

Now, let us consider a modeling of the classical two-slits experiment. Certainly, it is not possible to model infinite space in the real computer simulation. Thus, we are restricted by the consideration of the system in a box. However, due to good localization of $|\psi|^2$, box boundaries practically do not affect the simulation for a sufficiently large box and short times.

The evolution of the particle's wave function within a box will be described now by two-dimensional Schro\"dinger equation. For simplicity of numerical modeling, consider a half-width of the wave function crossing the slits,  $\sqrt{\alpha}$, as a unit length. This implies that $\tau=\hbar/2m\alpha$ is a unit time. Thus, dimensionless 2D Schr\"odinger equation takes the form:
\begin{equation}
i\frac{\partial \psi}{\partial t}=-\left(\frac{\partial^2 \psi}{\partial x^2}+\frac{\partial^2 \psi}{\partial y^2}\right).
\label{SchND}
\end{equation}

To visualize the solution, let us consider particle's motion in a space after slits. Since numerical simulation can deal with a finite region only, we consider a box with zero-fluxes boundary conditions on all its edges except the one, which contains the slits. For these slits we need to determine boundary conditions as symmetric (due to equivalence of both slits) influxes.

Since in the zero-time moment, particle flies into in a direction perpendicular to the wall, the influx boundary at the $x=0$ condition should connect normal spatial derivative of the wave function (\ref{wtrain}) with the wave function itself. In dimensionless form, it is written as
$$
\left.\frac{\partial \psi}{\partial x}\right|_{x=0}=
\left[i\frac{k_0}{1+t^2} +\frac{k_0 t}{1+t^2} \right]\psi.
$$

Finally, one need to use initial value for the wave function. In dimensionless form it has the  following form
\begin{equation}
\psi(x,y,0)=\frac{1}{2}\frac{1}{\sqrt[4]{2\pi}}e^{ik_0x}e^{-\frac{x^2}{4}},
\label{wtrainND}
\end{equation}
for $y$-s belong to slits and zero elsewhere.  

Results of simulations evaluated using {\sc FlexPDE} software are presented in the movie below. It demonstrates spatiotemporal evolution of the wave packet's squared amplitude.

\begin{center}
 \includemovie[
 poster,
]{0.5\linewidth}{0.38\linewidth}{psi2.mpeg}
\end{center}
{\small \sl Evolution of the probability density to detect particle after two slits. Red part of spectral colouring corresponds to larger values. [Click on the picture to see movie (Adobe Acrobat Reader required)].}

One can see that two initial bell-shaped curves with their centers located in the centers of slits are fast expanding. When the visible widths start to overlap each other, new maximum emerges between them. Thus, even small distance apart of initial slits, on most probably detects a particle in {\it three} places, not in two (see most bright red spots in the movie). Further, new maxima of probability density emerge in the transversal direction moving at the same time in the longitudinal one, i.e. developed interference pattern forms.

More detailed background for such a behaviour could be understood exploring spatio-temporal dynamics of the wave function itself. For example, the motions of real and imaginary parts of wave packets considered above are presented in the following movies.

\vspace{5mm}
\begin{minipage}{0.5\textwidth}
\begin{center}
 \includemovie[
 poster,
]{0.75\linewidth}{0.56\linewidth}{psir.mpeg}
\end{center}
\end{minipage}
\begin{minipage}{0.5\textwidth}
\begin{center}
 \includemovie[
 poster,
]{0.75\linewidth}{0.56\linewidth}{psii.mpeg}
\end{center}
\end{minipage}
{\small \sl Evolution of real (left) and imaginary (right) components of the wave function passed through two slits. Red part of spectral colouring corresponds to larger values. [Click on the picture to see movie (Adobe Acrobat Reader required)].}

\vspace{7mm}
One can see that each slit operates as a almost pointwise source of waves, which transforms with a time into concentric spherical ones like concentric overlapping waves on a water surface, which emerge if one periodically touches the surface in two points by fingers. Overlapping these concentric waves results in their interference. And the combination $|\psi|^2=\psi_r^2+\psi_i^2$ of these visible periodic patterns shifted on semiperiod in phase relatively to each other gives finally localized moving spots, which increase in number during the time.

\end{document}